\documentclass{aa}  
\usepackage{natbib}
\usepackage{graphicx}
\usepackage{txfonts}
\usepackage{lipsum}
\usepackage{subcaption}        
\usepackage{booktabs}
\usepackage{lscape}             
\usepackage{placeins}           
\usepackage{siunitx}
\usepackage{array}
\usepackage{float}
\usepackage{amsmath}
\usepackage{multirow}
\usepackage{xcolor}
\usepackage{fontawesome}
\usepackage{enumitem}
\usepackage{caption}
\usepackage[colorlinks=true,
linkcolor=blue,
citecolor=blue,
urlcolor=blue
]{hyperref}

\begin{document}
	
\title{Machine learning prediction of binary formation in three-body gravitational encounters}
	
\author{Ahmad Farhani Asl \inst{1}\corrauth{a.farhani@iasbs.ac.ir}}
	
\institute{
	Department of Physics, Institute for Advanced Studies in Basic Sciences (IASBS), Zanjan, Iran \\[0.5em]
	Secondary email: a.farhaniasl@gmail.com
}
	
\date{Received July 14, 2026}
	
\abstract
{Three-body encounters are frequent events in stellar systems, intrinsically chaotic, and computationally costly to model with direct $N$-body integration. Predicting whether such encounters lead to binary formation is therefore challenging, particularly in large-scale simulations.}
{We aim to develop an accurate and physically interpretable machine-learning model that predicts binary formation from the initial conditions of three-body encounters, to assess its reliability, and to identify the physical parameters that most strongly determine the outcome.}
{We trained an XGBoost binary classifier on a balanced dataset of three-body scattering experiments computed with the \texttt{REBOUND} code and the \texttt{IAS15} integrator. The input to the model consists of 30 physically motivated features describing the masses, energies, and kinematics of the initial configuration.}
{The classifier achieves excellent performance on a balanced test set, with accuracy, precision, recall, and F1-score all above 0.94, and with ROC--AUC and PR--AUC values of 0.99. Feature-importance analysis shows that the outcome is governed primarily by the mass hierarchy and hardness ratio of the encounter, followed by velocity fraction and mass entropy. The predicted probabilities are well calibrated, with an expected calibration error (ECE) of 0.02. Inference is approximately 400 times faster than direct N-body integration. The model also generalizes well across encounter radii, although its performance decreases in the weak-interaction regime where binary formation becomes intrinsically rare.}
{These results show that machine learning can provide fast, accurate, and physically interpretable predictions of binary formation in three-body encounters. Such models offer a practical complement to direct N-body simulations and may enable efficient probability estimates in large-scale simulations of stellar systems.}
	
	
\maketitle
	

\nolinenumbers
	
\section{Introduction}
	
Binaries are the simplest gravitationally bound stellar systems, yet their presence can govern the dynamical evolution of much larger structures such as star clusters and galactic nuclei. Two primary channels produce these binaries: primordial formation during the protostellar collapse phase \citep{Stacy2010, Offner2023}, or dynamical formation via few-body gravitational encounters \citep{Hut1985}. While two-body encounters have closed-form analytic solutions for both bound (Keplerian) and unbound (hyperbolic) orbits, a binary cannot form from an isolated two-body interaction because energy and angular momentum are conserved. A third body is required to carry away the excess kinetic energy, enabling the remaining two stars to become gravitationally bound. In any stellar environment, the probability of $N$ stars undergoing a simultaneous gravitational encounter decays sharply with $N$, making three-body encounters the most likely mechanism for dynamical binary formation.
	
The chaotic nature of the three-body problem \citep{Poincare1890, Lorenz1963, Valtonen2006} implies that initially similar configurations can diverge exponentially, leading to qualitatively different outcomes. Thus, whether a binary forms or not becomes inherently challenging to predict. Traditional approaches have relied on analytic scattering theories \citep[e.g.,][]{Hut1983} and extensive $N$-body simulation campaigns \citep{Heggie2003, Fregeau2006}. However, analytic methods typically provide cross-sections or probability distributions rather than individual outcome predictions. In contrast, direct $N$-body simulations, while accurate, are computationally expensive, and their cost rises steeply with increasing precision requirements \citep{Zwart2007, Yokota2012}.
	
Traditional approaches to three-body binary formation have recently seen significant extensions. \citet{Atallah2024} conducted direct numerical integrations of three initially unbound bodies, considering for the first time unequal masses over a broad parameter space. \citet{Ginat2024} developed an analytical probabilistic framework for a similar purpose, providing a closed-form expression for the binary formation probability based on conservation of energy and angular momentum. Despite these advances, predicting whether a given three-body encounter will result in binary formation directly from the initial conditions remains an open challenge.
	
Parallel advances in machine learning (ML) have opened new avenues for gravitational dynamics. \citet{Breen2020} demonstrates that deep neural networks can predict chaotic three-body trajectories up to 100 million times faster than numerical integrators, though their work focused on trajectories rather than binary formation outcomes. \citet{SiTu2025} extends this approach to post-Newtonian dynamics, achieving over 160 times acceleration while maintaining relative energy errors below 1\%. \citet{Pereira2024} introduces physics-informed neural networks (PINNs) that incorporate ordinary differential equations as regularizing agents, surpassing previous ML methods in prediction quality. \citet{Carita2024} applies image classification to identify retrograde resonances based on orbital shapes. Collectively, these studies establish that ML methods can capture complex gravitational interactions with speedups of several orders of magnitude. However, none have addressed the discrete outcome prediction of whether a three-body encounter forms a binary. 
	
To address this, we train an XGBoost ML classifier on three-body scattering simulations. From the initial conditions, we construct physically motivated features—mass ratios, velocity fractions, impact parameters, hardness, gravitational focusing, and angular momentum descriptors. The model predicts, as a binary classification, whether a given encounter forms a binary.
	
Our XGBoost classifier achieves near-perfect performance on a balanced test set, with accuracy, precision, recall, and F1-score all exceeding $0.94$ and both ROC-AUC and PR-AUC of $0.99$. The model is well-calibrated, with a Brier score of $0.0432$ and an expected calibration error (ECE) of $0.0202$. It also predicts at roughly $70{,}000$ samples per second, a speedup of approximately $400$ times over direct N-body integration. Feature importance ranks mass-related descriptors as dominant, followed by velocity-based quantities, with encounter geometry parameters playing a comparatively minor role. Performance remains robust across encounter radii (PR-AUC of $0.99$), with degradation only in the very weakly interacting regime, where the kinetic energy $U_\mathrm{kin}$ of the system significantly outweighs its potential energy magnitude $|U_\mathrm{pot}|$, and thus, binary formation is rare.
	
The remainder of this paper is organized as follows. Section~\ref{sec:method} describes the generation of the three-body scattering dataset and the ML methodology. Section~\ref{sec:results} presents the results, including model performance, feature importance, and speed benchmark. Section~\ref{sec:discussion} discusses the physical interpretation of the key findings and the limitations of our study. Finally, Section~\ref{sec:conclusions} concludes with a summary and an outlook for future work.
	
\section{Method}
\label{sec:method}
	
This section describes the methodology employed in this study. We first present the numerical setup for generating three-body encounter data, including the integration scheme, initial parameter sampling, and outcome classification. We then describe the construction of the physically motivated feature vector, the preprocessing steps, and the evaluation framework used to train and assess the ML model. 
	
\subsection{Three-Body Encounter Simulations}
	
We generated a synthetic dataset of three-body gravitational encounters using the REBOUND N-body package \citep{Rein2012}. For the integration, we used IAS15 \citep{Rein2015}, a 15th-order adaptive integrator that provides machine-precision energy conservation and robust handling of close encounters. Other REBOUND alternative integrators, while faster, are either lower in accuracy, e.g., the standard Wisdom-Holman symplectic integrator \citep{Wisdom1991}, or require parameter tuning through experimentation, e.g., the BS integrator \citep{Hairer1993}. Therefore, IAS15 is a well-suited integrator, particularly for the close encounters near the relativistic regime explored in this study. Furthermore, this integrator is widely adopted for similar purposes \citep{Dolcetta2020, Cesare2021, Tory2022}.
	
All simulations were performed using $G = 4\pi^2$ in units of years (yr), astronomical units (au), and solar masses (M$_\odot$). We assumed point-mass bodies interacting within the Newtonian regime. To apply the latter restriction, we excluded simulations in which bodies approached closer than $50$ times their Schwarzschild radius. Integrations terminated when any body exceeded a distance of $10^7$ au from the center of mass or the simulation time reached $t_{\max}=3\,t_{\mathrm{enc}}$, where $t_{\mathrm{enc}}$ is the approximate time taken for the bodies to reach the encounter region (see Appendix~\ref{app:sample_visual} for example visualizations). 
	
\subsubsection{Initial Conditions}
\label{sec:sampling}
All simulations satisfy the initially unbound condition, in which hardness $\mathcal{H} = |U_\mathrm{pot}|/U_\mathrm{kin} \in [0, 1)$, where $U_\mathrm{pot}$ and $U_\mathrm{kin}$ are potential and kinetic energies, respectively. Random generation exhibits a strong bias toward the highly unbound regime ($\mathcal{H}$ much less than one), where the phase space volume is largest. Configurations with $\mathcal{H}$ close to one, by contrast, occupy a significantly smaller phase space volume and are consequently severely undersampled. To obtain a balanced sample across the full unbound range, we performed stratified sampling in hardness bins, targeting nearly equal numbers of simulations per bin.
	
\begin{figure}[t!]
	\centering
	\includegraphics[width=\linewidth]{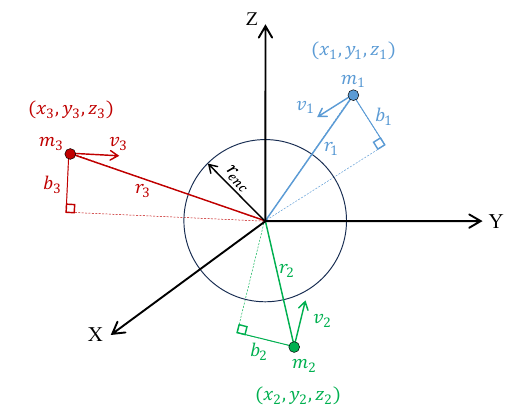}
	\caption{Initial configuration for a three-body encounter. Each star $i$ ($i = 1,2,3$) has mass $m_i$, impact parameter ${b}_i$, initial velocity $v_i$, and initial position ${r}_i$. The encounter radius $r_{\mathrm{enc}}$ defines the interaction volume scale.}
	\label{fig:initial_config}
\end{figure}
	
\begin{table}[t!]
	\centering
	\caption{Initial parameter ranges used in the three-body encounter simulations.}
	\label{tab:initial_params}
	\begin{tabular}{p{3.2cm} p{1.5cm} l}
		\toprule
		Parameter & Symbol & Range \\
		\midrule
		Stellar masses & $m_i$ & $0.08$--$150$ $M_\odot$ \\
		Initial velocities & $v_i$ & $0.01$--$100$ au yr$^{-1}$ \\
		Encounter radius & $r_{\mathrm{enc}}$ & $0.01$--$10000$ au \\
		Impact parameters & $b_i$ & $[0, r_{\mathrm{enc}}]$ \\
		Polar angle & $\theta$ & $\cos\theta \in [-1, 1]$  \\
		Azimuthal angle & $\phi$ & $[0, 2\pi]$ \\
		Hardness & $\mathcal{H}$ & $[0, 1)$ \\
		\bottomrule
	\end{tabular}
\end{table}
	
As shown in Table~\ref{tab:initial_params}, each simulation sampled three stellar masses from a log-uniform distribution between $0.08$ and $150 M_\odot$ \citep{Kroupa2001, Weidner2004}. Initial velocities were drawn from a uniform distribution between $10^{-2}$ and $10^2$ au yr$^{-1}$, corresponding to velocity dispersions from roughly 0.05 to 500 km s$^{-1}$, spanning open clusters to galactic nuclei \citep{Binney2008, Bojnordi2021}. The angles $\theta$ (polar) and $\phi$ (azimuthal) were chosen isotropically: $\cos\theta$ uniform in $[-1,1]$, $\phi$ uniform in $[0,2\pi)$. The encounter scale was controlled by the parameter $r_{\mathrm{enc}}$, sampled log-uniformly between $0.01$ and $10^4$ au. This range corresponds to the interaction scales associated with stellar number densities spanning from approximately 0.1 pc$^{-3}$ (open clusters) to 10$^8$ pc$^{-3}$ (ultra-dense star-forming cores), via the relation $n \propto r_{\mathrm{enc}}^{-3}$ \citep{Davies1996, Torres2013}. Hence, the closest approach distance was set to $r_{\min} = 10 r_{\mathrm{enc}}$, and the encounter time was defined as $t_{\mathrm{enc}} = r_{\min} / v_{\min}$, where $v_{\min}$ is the smallest initial velocity among the three bodies. The initial position and velocity of star $i$ were then calculated as
\[
	\mathbf{r}_i = v_i t_{\mathrm{enc}} \hat{\mathbf{v}}_i + \mathbf{b}_i,\qquad
	\mathbf{v}_i = -v_i \hat{\mathbf{v}}_i,
\]
where $\hat{\mathbf{v}}_i$ and $\mathbf{b}_i$ are the approach direction unit vector, determined by ($\theta$, $\phi$), and impact parameter vector, respectively. This construction ensures all three bodies reach the encounter region at approximately the same time $t_{\mathrm{enc}}$. Figure~\ref{fig:initial_config} illustrates the initial configuration geometry.
	
\subsubsection{Binary Identification}
	
Since the total energy is initially positive (unbound system), formation of a bound hierarchical triple is very unlikely (requires $E_\mathrm{tot}<0$). We identify the binary as the pair with the lowest (most negative) pairwise energy upon reaching the final integration time $t_\mathrm{max}$
\[
	E_{ij} = \frac{1}{2} \mu_{ij} |\mathbf{v}_i - \mathbf{v}_j|^2 - \frac{G m_i m_j}{|\mathbf{r}_i - \mathbf{r}_j|},
\]
with $\mu_{ij}=m_i m_j/(m_i+m_j)$ the reduced mass. If no pair had $E_{ij}<0$, the simulation was classified as no binary formation. For simulations with a formed binary, we additionally verified that the third body was unbound from the binary. To ensure that the detected binaries are non-transient, we extended the integration time up to \(10\ t_{\mathrm{max}}\) for cases with binary formation. For each formed binary, we recorded its orbital parameters (see Appendix~\ref{app:orbital}).
	
\subsection{Machine Learning Model}
	
We chose to train an XGBoost classifier \citep{Chen2016}, as gradient boosting algorithms have consistently demonstrated state-of-the-art performance on tabular data, often outperforming deep learning models in classification tasks \citep{Albertyn2025}. We selected hyperparameters through a combination of automated optimization and manual exploration. We first performed random search using scikit-learn's built-in function \citep{Bergstra2012} to identify a promising parameter range. We then manually explored refinements around the best-performing configurations, tracking performance metrics iteratively. The final parameters that yielded the best performance are 1400 estimators, a maximum depth of 18, a learning rate of 0.02, subsample and column sample fractions of 0.9 and 1.0, and L2 regularization \(\lambda = 1.0\). We used GPU acceleration for training and evaluated generalization via 10-fold stratified cross-validation using the average precision (PR-AUC) score.
	
\subsubsection{Feature Construction}
	
From each simulated three-body encounter, we constructed 30 features based on the initial conditions. The three bodies are ordered by their masses, with the indices minus ($-$), circle ($\circ$), and plus ($+$) denoting the least, intermediate, and most massive star, respectively. These indices are carried through all features and refer to the star by its mass rank (do not confuse with the minimum, median, or maximum value of the feature itself). Table~\ref{tab:features} lists all features with their symbols and definitions.
	
\begin{table}[t!]
	\centering
	\caption{Constructed features for binary formation prediction. Indices $-$, $\circ$, and $+$ correspond to least, intermediate, and most massive stars, respectively.}
	\label{tab:features}
	\begin{tabular}{p{4.5cm} l}
		\toprule
		Feature & Description \\
		\midrule
		\makebox[0.7cm][l]{$m_i$} masses &
		$m_\mathrm{-}$, $m_\mathrm{\circ}$, $m_\mathrm{+}$ \\		
		\makebox[0.7cm][l]{$M$} total mass &
		$\sum_i m_i$ \\		
		\makebox[0.7cm][l]{$\mu_i$} mass fractions &
		$m_i/M$ \\		
		\makebox[0.7cm][l]{$\mu_{\pm}$} mass hierarchy &
		$m_{+}/m_{-}$ \\		
		\makebox[0.7cm][l]{$\mu_p$} mass pairs summation &
		$\sum_i\sum_{j>i} m_i m_j$ \\		
		\makebox[0.7cm][l]{$\mu_c$} mass concentration &
		$(m_{\circ}+m_{-})/m_{+}$ \\		
		\makebox[0.7cm][l]{$\mu_a$} mass asymmetry &
		$(m_{+}-m_{-})/(m_{+}+m_{-})$ \\		
		\makebox[0.7cm][l]{$\delta_m$} mass entropy &
		$\left|\sum_i (m_i/M)\ln(m_i/M)\right|$ \\		
		\makebox[0.7cm][l]{$v_i$} velocities &
		$v_\mathrm{-}$, $v_\mathrm{\circ}$, $v_\mathrm{+}$ \\		
		\makebox[0.7cm][l]{$\upsilon_i$} velocity fractions &
		$v_i/v_{\rm esc}$ \\		
		\makebox[0.7cm][l]{$\upsilon_{\pm}$} velocity hierarchy &
		$v_{+}/v_{-}$ \\		
		\makebox[0.7cm][l]{$\upsilon_a$} velocity asymmetry &
		$|v_{+}-v_{-}|/(v_{+}+v_{-})$ \\		
		\makebox[0.7cm][l]{$\bar{\upsilon}$} mean escape velocity &
		$\sqrt{2GM/r_{\rm enc}}\,/\,\bar v$ \\		
		\makebox[0.7cm][l]{$b_i$} impact parameters &
		$b_\mathrm{-}$, $b_\mathrm{\circ}$, $b_\mathrm{+}$ \\		
		\makebox[0.7cm][l]{$\psi_{ij}$} relative directional angles &
		$\psi_\mathrm{-\circ}$, $\psi_\mathrm{-+}$, $\psi_\mathrm{\circ+}$ \\		
		\makebox[0.7cm][l]{$\mathcal{F}$} gravitational focusing &
		$GM/(\bar b\,\bar v^2)$ \\		
		\makebox[0.7cm][l]{$\mathcal{R}$} virial radius cutoff &
		$3r_{\rm enc}/(G\mu_p)$ \\		
		\makebox[0.7cm][l]{$\mathcal{H}$} hardness &
		$|U_{\rm pot}|/U_{\rm kin}$ \\		
		\makebox[0.7cm][l]{$\ell_{\rm bin}$} scaled angular momentum &
		$L/U_{\rm bin}$ \\
		\bottomrule
	\end{tabular}
\end{table}
	
The feature set was designed to capture the physical parameters governing three-body binary formation. Mass-based features encode the role of mass distribution across three stars that determine the efficiency of energy exchange during close encounters \citep{Heggie1996}. Velocity-based features quantify the competition between kinetic energy and gravitational binding, distinguishing hard from soft encounter regimes \citep{Trani2019}. Impact parameters $b_i$ set the geometric scale of each star's approach, directly influencing gravitational focusing \citep{Valtonen2006}. The relative directional angles $\psi_{ij}$ between velocity vectors capture the orientation of the encounter, which affects angular momentum transport and the likelihood of resonant interactions \citep{Hall1996}.
	
We also included several composite dynamical features. The gravitational focusing factor $\mathcal{F} = GM/(\bar{b}\bar{v}^2)$ is a standard measure of the enhancement of the encounter cross-section due to gravity. To encapsulate a numerical scaling relation for binary formation probability from three initially unbound bodies, we adopt a virial-radius cutoff $\mathcal{R} = 3r_{\mathrm{enc}}/(G\mu_p)$ for the interaction volume, inspired by the energy-based cutoff definition used in the analytical work of \citet{Ginat2024}. The hardness ratio $\mathcal{H} = |U_{\mathrm{pot}}|/U_{\mathrm{kin}}$ is the fundamental parameter distinguishing bound from unbound systems; within the unbound regime ($\mathcal{H}<1$), it correlates with the likelihood of binary formation \citep{Heggie1975, Trani2019}. Finally, the scaled angular momentum $\ell_{\mathrm{bin}} = L/|U_{\mathrm{bin}}|$ determines the final binary eccentricity distribution by enforcing how the conserved total angular momentum is partitioned between the binary orbit and the escaping star \citep{Valtonen2006}.
	
\subsubsection{Data Preprocessing}
\label{sec:preprocessing}
	
Our raw dataset comprised $359,149$ simulated encounters, of which $118,076$ (approximately $33\%$) resulted in binary formation. To ensure that the retained cases are physically reliable, we applied four quality cuts. First, we discarded simulations with a relative energy error $|\Delta E/E|$ greater than $10^{-8}$, which removed $4\%$ of the entire raw dataset. This threshold is consistent with high-accuracy integration standards in state-of-the-art N-body codes such as PeTar \citep{Wang2020} and is considerably more stringent than the roughly $10^{-1}$ level sufficient for ensemble-level statistics of chaotic three-body scattering \citep{Zwart2014}. Second, for systems that formed a binary, we retained only those with semi-major axis $a$ smaller than $10^5$ au, excluding $4.7\%$ of the binaries. This limit is motivated by Gaia-based observations of the field wide-binary population \citep{Makarov2025}. Third, we excluded binaries with orbital eccentricity exceeding $e = 0.9999$, which removed an additional $13.8\%$ of the binaries. Numerical studies have shown that this threshold represents a near-parabolic worst-case regime where standard solvers of Kepler's equation become ill-conditioned and prone to significant accuracy degradation \citep{Mortari2007}. Fourth, we required that binaries complete at least two orbital periods within the extended integration time, ensuring that the binary is dynamically established and its properties are not affected by the finite simulation time. This criterion excluded $25\%$ of the raw binary-forming cases, predominantly those with extremely long-period orbits or formations much later than the encounter time $t_\mathrm{enc}$. Additionally, during the simulation process, $3.5\%$ of binary-forming cases were excluded as transient binaries (i.e., binaries that did not survive over the extended integration). Therefore, the final cleaned dataset comprised $312,666$ simulations, with a class distribution of $26\%$ positive (binary formation) and $74\%$ negative (no binary formation). Readers may notice an apparent inconsistency between the number of removed cases and the final class distribution; this arises from significant overlap among the quality cuts. 
	
Since our cleaned dataset is inherently imbalanced, machine learning classifiers can be biased toward the majority (negative) class, yielding poor sensitivity in predicting the minority class of binary formation. To mitigate this, following \citet{Chen2004}, we randomly undersampled the majority class to create a balanced dataset with 1:1 class ratio. Random undersampling is a widely adopted strategy for imbalanced binary classification in astrophysical machine-learning applications, with similar balancing procedures employed in studies such as radio galaxy cross-identification and pulsar detection \citep{Alegre2022, Lin2020}. Given that the positive class contains over $10^{5}$ samples, the dataset remains sufficiently large after undersampling, allowing the majority class to be reduced without significant information loss. We additionally tested class-weighting \citep{Wainer2018}, a cost-sensitive approach that assigns higher misclassification costs to the minority class, thereby forcing the model to focus on learning from underrepresented instances \citep{Sasirekha2025}. However, we found that undersampling yielded consistently superior performance across all metrics and was therefore adopted for our primary analysis. 
	
The balanced dataset was subsequently divided into training and test sets using an 80/20 split. This ratio is widely adopted in machine-learning applications \citep{Gholamy2018} and provides a good compromise between maximizing the amount of training data and retaining a sufficiently large independent test set. We additionally verified that alternative 90/10 and 70/30 splits produced results consistent with the 80/20 split, with differences of at most $0.5\%$ in all metrics. No feature scaling was applied, as tree-based ensemble methods such as XGBoost remain robust regardless of scaling \citep{Pinheiro2025}.
	
\subsubsection{Evaluation Metrics}
	
Model performance was assessed using accuracy, precision, recall, F1-score, ROC-AUC, PR-AUC, Brier score, and expected calibration error (ECE). These metrics were computed using scikit-learn \citep{Pedregosa2011}. Confusion matrices were visualized to inspect classification behavior, and performance was analyzed as a function of encounter radius $r_{\mathrm{enc}}$ and hardness $\mathcal{H}$ to identify regions of strength and weakness. Physical baselines, derived from the most important features, were established for benchmarking the XGBoost model. Finally, the model was evaluated on a new, randomly generated subset of simulated data to test its generalization performance on a more realistic and extremely imbalanced data distribution.
	
\section{Results}
\label{sec:results}
	
In this section, we report the results of our trained XGBoost classifier. We first present the overall performance metrics on the held-out test set, followed by an analysis of model performance as a function of encounter radius and hardness. We then examine the feature importance ranking, compare the model against physically motivated baselines, and benchmark its inference speed against direct N-body integration.
	
\subsection{Model Performance}
	
We evaluated the XGBoost classifier on the held-out test set (approximately $32{,}500$ samples), where the model achieves above $0.94$ for accuracy, precision, recall, and F1-score. Both ROC-AUC and PR-AUC are nearly $0.99$. The Brier score and expected calibration error (ECE) are $0.0432$ and $0.0202$, respectively.
	
\begin{figure}[t!]
	\centering
	\includegraphics[width=\columnwidth]{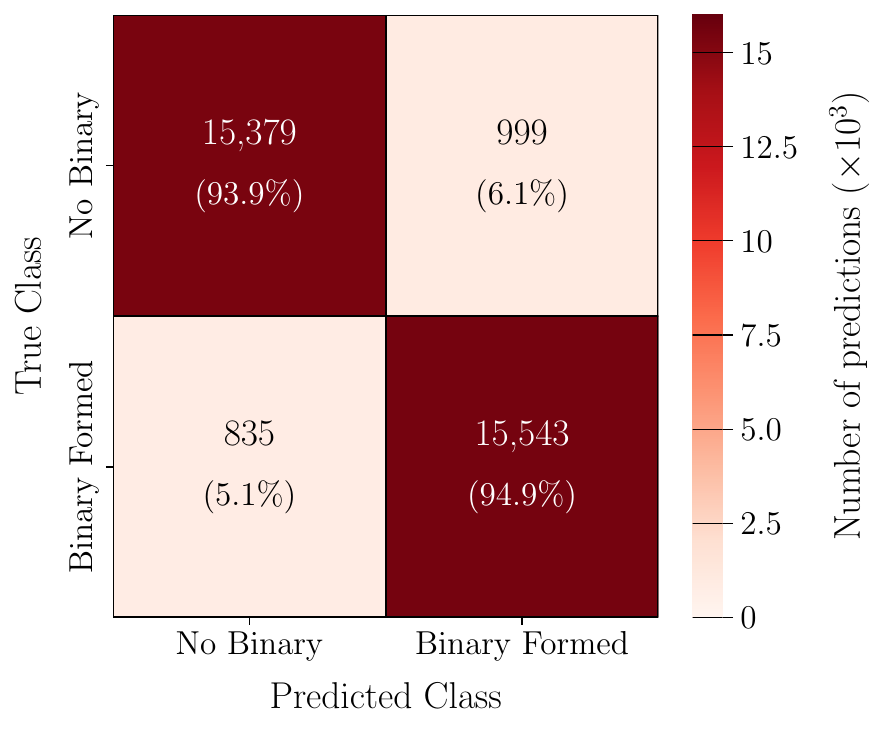}
	\caption{Confusion matrix evaluated on the test set.}
	\label{fig:confusion_matrix}
\end{figure}

The confusion matrix (Figure~\ref{fig:confusion_matrix}) shows $15,379$ true negatives, $835$ false positives, $999$ false negatives, and $15,543$ true positives. Ten-fold stratified cross-validation on the training set yields a mean PR-AUC of $0.9853\pm 0.0008$.
	
\subsection{Performance Across Parameter Space}
	
Figure~\ref{fig:across_performance} summarizes model performance as a function of the encounter radius $r_{\mathrm{enc}}$ (left panel) and hardness $\mathcal{H}$ (right panel). In both panels, the green bars show the binary fraction ($N_b / N_{\mathrm{tot}}$, where $N_b$ is the number of simulations resulting in binary formation and $N_{\mathrm{tot}}$ is the total number of simulations) in the test set, and the red squares with connecting lines show the achieved PR-AUC. For $r_{\mathrm{enc}}$, the binary fraction decreases monotonically from $0.63$ for $r_{\mathrm{enc}} \in [0.01, 1)$ au to $0.37$ for $r_{\mathrm{enc}} \in [100, 10{,}000)$ au. The PR-AUC remains high, approximately $0.99$, across all bins. For hardness, the binary fraction increases sharply with $\mathcal{H}$, from $0.09$ for $\mathcal{H} \in [0, 0.01)$ to $0.64$ for $\mathcal{H} \in [0.1, 1.0)$. The PR-AUC rises correspondingly, from $0.73$ in the lowest hardness bin to $0.99$ in the highest. 
	
\subsection{Feature Importance}
	
\begin{figure}[b!]
	\centering
	\includegraphics[width=0.95\columnwidth]{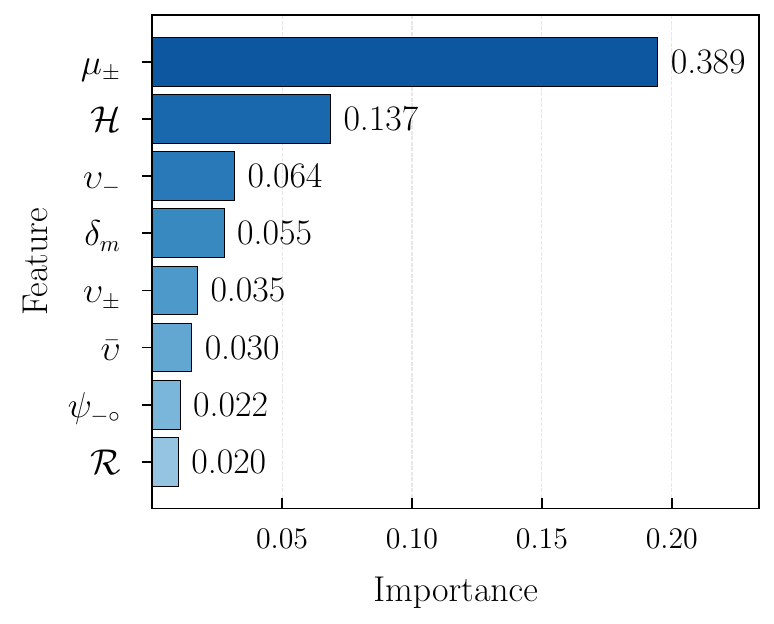}
	\caption{Top important features of the XGBoost classifier.}
	\label{fig:feature_importance}
\end{figure}

Figure~\ref{fig:feature_importance} displays the most influential features of the trained model. The mass hierarchy $\mu_\pm$ is the most important feature, with an importance of nearly $0.4$, roughly three times higher than that of the second most important feature, hardness $\mathcal{H}$ (with importance of $0.14$). The velocity fraction $\upsilon_-$, mass entropy $\delta_{m}$, and velocity hierarchy $\upsilon_{\pm}$ are also ranked high, though their importance gradually drops from $0.065$ to $0.035$. All remaining features, including the relative directional angle $\psi_{-\circ}$, virial-radius cutoff $\mathcal{R}$, and gravitational focusing $\mathcal{F}$, have importance scores below $0.03$. 
	
\subsection{Model Comparison and Inference Speed}
	
The best-performing physical baselines are mass hierarchy ($\mu_\mathrm{\pm}$) and mass asymmetry ($\mu_{a}$) from the training features (see Table~\ref{tab:features}), each achieving a PR-AUC of $0.72$, while all other single-feature baselines fall below $0.60$. In contrast, our trained model achieves $0.99$.
	
We benchmarked the inference speed of the trained XGBoost model on approximately $800{,}000$ new simulated three-body encounters. We generated this dataset from fully random initial parameter distributions, unlike the stratified sampling approach used for producing the training data (see Section~\ref{sec:sampling}). We preprocessed this new dataset (see Section~\ref{sec:preprocessing}) and neither used it for training nor for evaluation. The model achieves about $70{,}000$ predictions per second. For comparison, the IAS15 integrator processes approximately 165 simulations per second on the same hardware, an Intel Core i7-7700K CPU running at 4.2 GHz. Thus, the XGBoost model provides a speedup factor of around $400$ relative to direct N-body integration.
	
Table~\ref{tab:speed_evaluation} summarizes the model's performance and inference speed on both the full raw dataset and the balanced subset (produced by randomly undersampling the majority class). The raw dataset is extremely imbalanced, with only $0.05$ of cases forming a binary. Recall ($0.65$) and ROC-AUC ($0.96$) are identical across both datasets. Precision, F1-score, and PR-AUC are lower on the raw dataset. Nevertheless, the model's performance on this highly imbalanced dataset remains comparable to the best physical baseline trained on a perfectly balanced dataset, and substantially exceeds random guessing, which has only a $0.05$ chance of detecting a binary formation.  
		
\begin{figure}[t!]
	\centering
	\includegraphics[width=\linewidth]{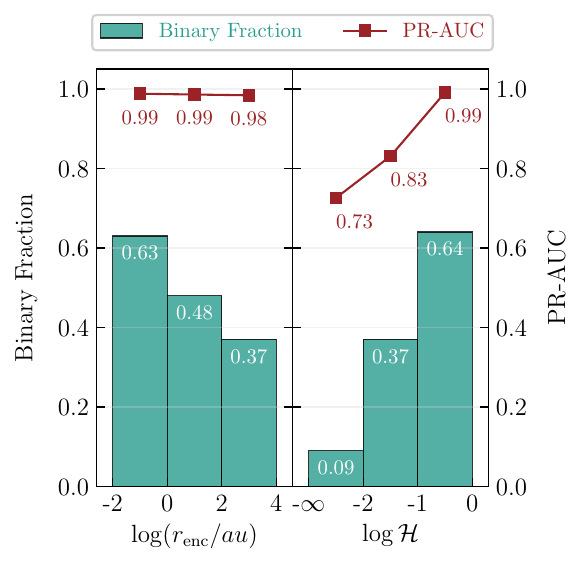}
	\caption{Model performance versus encounter radius $r_{\mathrm{enc}}$ (left) and hardness $\mathcal{H}$ (right). Green bars indicate the binary fraction in each range of the test set. Red squares with connecting lines indicate the model's achieved PR-AUC over each range.}
	\label{fig:across_performance}
\end{figure}
	
\begin{table}[h!]
	\centering
	\caption{Model performance and inference speed on the full raw dataset and the balanced subset.}
	\label{tab:speed_evaluation}
	\begin{tabular}{p{2.7cm}p{2cm}l}
		\toprule
		Metric & Raw dataset & Balanced dataset \\
		\midrule
		Samples & 827,000 & 88,750 \\
		Binary fraction & 0.05 & 0.5 \\
		\midrule
		Predictions (s$^{-1}$) & 63,973 & 68,494 \\
		Speedup factor & 376$\times$ & 403$\times$ \\
		\midrule
		Accuracy & 0.95 & 0.82 \\
		Precision & 0.56 & 0.96 \\
		Recall & 0.65 & 0.66 \\
		F1-score & 0.60 & 0.78 \\
		ROC-AUC & 0.96 & 0.96 \\
		PR-AUC & 0.66 & 0.96 \\
		\bottomrule
	\end{tabular}
\end{table}
	
\section{Discussion}
\label{sec:discussion}
	
Our trained XGBoost model, with a PR-AUC of 0.99, excels in identifying binary-forming three-body encounters, with only about $5\%$ false predictions. According to the 10-fold cross-validation results, the model generalizes well without overfitting. The Brier score and expected calibration error (ECE) suggest adequate probability calibration, making the model suitable for probabilistic applications such as Monte Carlo sampling in simulating dense star clusters \citep{Rodriguez2022}.   
	
For $r_{\mathrm{enc}}$, the binary fraction decreases with increasing encounter radius, consistent with weaker gravitational interactions at larger scales. The PR-AUC remains high (around $0.99$) across all bins, demonstrating robust generalization across diverse astronomical environments, particularly Galactic star clusters \citep{Atallah2025}.
	
At the lowest hardness bin ($\mathcal{H}$ less than $0.01$), where the model struggles the most with a PR-AUC of $0.7$, the binary fraction is as low as $0.1$. Such severe class imbalance biases the model toward the majority class, reducing sensitivity to rare events \citep{Nawaz2025}. Since PR-AUC directly evaluates performance on the minority class, this metric drops considerably when datasets are significantly imbalanced \citep{Mojiri2021}. Nevertheless, the model still substantially outperforms random guessing, indicating that the XGBoost classifier learns relevant physics beyond mere statistics. Class imbalance, however, is not the only factor governing performance. As shown in Figure~\ref{fig:across_performance}, two subsets with identical binary fraction ($0.37$) yield markedly different PR-AUC scores: $0.98$ for $r_{\mathrm{enc}}$ and $0.83$ for $\mathcal{H}$. This discrepancy suggests that the intrinsic separability of the feature space differs between the two parameters, with hardness likely producing a more ambiguous decision boundary than encounter radius at the same class ratio.      
	
The dominance of mass hierarchy ($\mu_\pm$) is consistent with mass distribution governing pairing probabilities in three-body encounters \citep{Atallah2024}. Hardness ($\mathcal{H}$), the second most important feature, directly probes the energy budget of the encounter, quantifying whether the system is energetically favorable for binary formation. Its importance is therefore physically expected, as it encodes the fundamental competition between binding and kinetic energy \citep{Heggie1975}. The remaining top-ranking features — velocity descriptors, mass entropy, directional angles, and the virial-radius cutoff — collectively reflect the model's sensitivity to kinetic energy partitioning, mass diversity, and encounter geometry, though their individual contributions are comparatively modest \citep{Valtonen2006, Samsing2018, Ginat2024}. Unlike black-box deep learning models, XGBoost's feature importance provides direct physical interpretability, making it a valuable tool for scientific discovery in gravitational dynamics \citep[see][]{Pasquato2024, Valle2021}.
	
The large gap between the best physical baseline (PR-AUC $0.72$) and XGBoost trained model (PR-AUC $0.99$) demonstrates that predicting binary formation requires nonlinear combinations of features beyond simple threshold-based criteria. The speedup of approximately 400 times over REBOUND IAS15 direct integration makes the model suitable for rapid, large-scale binary population synthesis \citep{Ivanova2005, Marks2012}. The identical recall and ROC-AUC across raw and balanced datasets confirm that the model's ability to identify true binary-forming events, which is the main goal of this study, is unaffected by class imbalance. The lower precision on the raw dataset is expected, as precision is inherently sensitive to the number of negative samples; in a highly imbalanced setting, even a small number of false positives leads to a substantial drop in precision. Accordingly, the lower PR-AUC on the raw data reflects the same effect, as PR-AUC is computed over precision-recall trade-offs.
	
Several limitations should be acknowledged. Our simulations assume point-mass bodies and exclude post-Newtonian corrections and physical collisions, restricting the model's applicability to nonrelativistic, nonmerging encounters. The hardness range is limited to $\mathcal{H} \in [0, 0.9)$; configurations with $\mathcal{H}$ very close to one (or greater than $0.9$) are extremely rare in random sampling and could not be generated in sufficient numbers. Furthermore, the model was trained on isolated three-body encounters, neglecting the background tidal field and external perturbations significantly present in dense stellar environments such as galactic nuclei and globular clusters. Finally, the initial parameter distributions were drawn from a log-uniform random sampling, which may not fully represent the phase-space distributions of realistic stellar populations with specific mass functions and velocity dispersions. 
	
\section{Summary and Conclusions}
\label{sec:conclusions}
	
In this work, we developed an XGBoost classifier to predict binary formation from initial conditions in three-body encounters—a task that has traditionally required expensive numerical integrations. We generated a synthetic dataset of three-body gravitational encounters, all initially unbound over a comprehensive range of initial parameters corresponding to different stellar environments, using REBOUND with the IAS15 integrator. From each simulation, we constructed physically motivated features. The model was trained and evaluated on a balanced dataset.
	
Our main findings are as follows:
	
\begin{enumerate}

	\item The XGBoost classifier achieves near-perfect performance on the balanced test set, with all metrics exceeding $0.94$ and ROC-AUC and PR-AUC of $0.99$. Ten-fold cross-validation confirms strong generalization ($0.9853 \pm 0.0008$). The model is well-calibrated (Brier $0.0432$, ECE $0.02$).\\
		
	\item Mass hierarchy ($\mu_\pm$) and hardness ($\mathcal{H}$) are the dominant predictors, followed by velocity fraction ($\upsilon_-$), and mass entropy ($\delta_m$). This ranking is broadly consistent with theoretical expectations, while the relatively lower importance of geometric features suggests that the initial encounter geometry plays a secondary role in determining binary formation outcomes.\\
		
	\item The model generalizes robustly across encounter radii, maintaining a PR-AUC around $0.99$ across all $r_{\mathrm{enc}}$ bins. Performance degrades (a PR-AUC of $0.73$) only in the very weakly interacting regime ($\mathcal{H}$ much less than $0.01$), where the binary fraction is under $0.10$; this is attributed to severe class imbalance rather than model failure and confirms that the model learns relevant physics rather than relying merely on statistics.\\
		
	\item The model outperforms the best single-feature physical baseline by a substantial margin (PR-AUC $0.99$ vs. $0.72$), demonstrating that binary formation requires nonlinear combinations of features beyond simple thresholds.\\
		
	\item With an inference speed of roughly $70{,}000$ predictions per second and a speedup of $400$ times over direct N-body integration, the model is well-suited for rapid probability estimation in large-scale cluster simulations and population synthesis.
\end{enumerate}
	
Future work will extend this approach to include post-Newtonian corrections and collisions, broadening its applicability to compact object encounters relevant to gravitational wave astrophysics. Incorporating additional features, such as initial relative phases and angular momentum distributions, could further improve performance in the low-hardness regime. Applying the model to large-scale cluster simulations via Monte Carlo methods would provide a direct test of its practical utility, addressing the longstanding gap between direct N-body integration and faster but less accurate prescriptions \citep{Atallah2024, Ginat2024}. Additionally, embedding the model as a surrogate within established N-body codes such as \texttt{NBODY6++} \citep{Aarseth1999} would enable direct validation in realistic cluster evolution scenarios and facilitate its adoption by the broader astrophysical community.
	
This work demonstrates that ML can complement classical three-body theory by providing fast, accurate, and interpretable predictions of encounter outcomes. The feature importance analysis offers physical insight into the factors governing binary formation, while the model's speed enables rapid probability estimation in simulations where resolving every encounter is computationally prohibitive. By bridging the gap between expensive N-body integrations and fast analytical approximations, this approach opens new avenues for modeling binary formation in dense stellar systems.
	
\section{Data Availability}
	
The simulation data (initial conditions, features, and outcomes) and the trained XGBoost model presented in this work will be deposited at the CDS (Centre de Données astronomiques de Strasbourg) upon acceptance of the paper. The complete code used to generate the data and train the model is publicly available on GitHub at \hyperref{https://github.com/ahmed-farhani-asl/Machine-learning-prediction-of-binary-formation-in-three-body-gravitational-encounters}{}{}{\faGithub}.
	
\section*{Acknowledgements}
	
We acknowledge the use of ChatGPT (OpenAI) and DeepSeek (DeepSeek AI) for language refinement, editing assistance, code debugging, and literature search during the preparation of this manuscript. These tools were used solely to improve clarity and readability, assist with code development, and retrieve relevant references. All AI-assisted outputs were reviewed, verified, and controlled by the author. All scientific content, analyses, results, interpretations, and final code are the sole responsibility of the author.
	
\bibpunct{(}{)}{;}{a}{}{,} 
\bibliographystyle{aa}
\bibliography{references}

@ARTICLE{Heggie1975,
	author = {{Heggie}, D.~C.},
	title = "{Binary evolution in stellar dynamics.}",
	journal = {\mnras},
	year = 1975,
	month = dec,
	volume = {173},
	pages = {729-787},
	doi = {10.1093/mnras/173.3.729},
	adsurl = {https://ui.adsabs.harvard.edu/abs/1975MNRAS.173..729H}
}

@article{Hut1985,
	author = {{Hut}, P.},
	title = "{Binary formation and interactions with field stars}",
	journal = {IAU Symposium},
	volume = {113},
	pages = {231-236},
	year = {1985},
	adsurl = {https://ui.adsabs.harvard.edu/abs/1985IAUS..113..231H},
	note = {Binary-single star scattering experiments}
}

@INPROCEEDINGS{Offner2023,
	author = {{Offner}, S.~S.~R. and {Moe}, M. and {Kratter}, K.~M. and {Sadavoy}, S.~I. and {Jensen}, E.~L.~N. and {Tobin}, J.~J.},
	title = "{The Origin and Evolution of Multiple Star Systems}",
	booktitle = {Protostars and Planets VII},
	year = 2023,
	editor = {{Inutsuka}, S. and {Aikawa}, Y. and {Muto}, T. and {Tomida}, K. and {Tamura}, M.},
	series = {Astronomical Society of the Pacific Conference Series},
	volume = {534},
	month = jul,
	pages = {275},
	doi = {10.48550/arXiv.2203.10066},
	archivePrefix = {arXiv},
	eprint = {2203.10066},
	primaryClass = {astro-ph.SR},
	adsurl = {https://ui.adsabs.harvard.edu/abs/2023ASPC..534..275O}
}

@ARTICLE{Atallah2024,
	author = {{Atallah}, Dany and {Weatherford}, Newlin C. and {Trani}, Alessandro A. and {Rasio}, Frederic A.},
	title = "{On Binary Formation from Three Initially Unbound Bodies}",
	journal = {\apj},
	year = 2024,
	month = aug,
	volume = {970},
	number = {2},
	eid = {112},
	pages = {112},
	doi = {10.3847/1538-4357/ad5185},
	archivePrefix = {arXiv},
	eprint = {2402.12429},
	primaryClass = {astro-ph.SR},
	adsurl = {https://ui.adsabs.harvard.edu/abs/2024ApJ...970..112A}
}

@ARTICLE{Ginat2024,
	author = {{Ginat}, Yonadav Barry and {Perets}, Hagai B.},
	title = "{Three-body binary formation in clusters: analytical theory}",
	journal = {\mnras},
	year = 2024,
	month = jun,
	volume = {531},
	number = {1},
	pages = {739-750},
	doi = {10.1093/mnras/stae1241},
	archivePrefix = {arXiv},
	eprint = {2404.08040},
	primaryClass = {astro-ph.GA},
	adsurl = {https://ui.adsabs.harvard.edu/abs/2024MNRAS.531..739G}
}

@ARTICLE{Breen2020,
	author = {{Breen}, Philip G. and {Foley}, Christopher N. and {Boekholt}, Tjarda and {Portegies Zwart}, Simon},
	title = "{Newton versus the machine: solving the chaotic three-body problem using deep neural networks}",
	journal = {\mnras},
	year = 2020,
	month = may,
	volume = {494},
	number = {2},
	pages = {2465-2470},
	doi = {10.1093/mnras/staa713},
	archivePrefix = {arXiv},
	eprint = {1910.07291},
	primaryClass = {astro-ph.GA},
	adsurl = {https://ui.adsabs.harvard.edu/abs/2020MNRAS.494.2465B}
}

@article{Poincare1890,
	author = {Poincaré, Henri},
	title = {Sur le problème des trois corps et les équations de la dynamique},
	journal = {Acta Mathematica},
	volume = {13},
	pages = {1--270},
	year = {1890},
	doi = {10.1007/BF02392506},
	url = {https://doi.org/10.1007/BF02392506},
	note = {Oeuvres, tome VII, pp. 262-479}
}

@ARTICLE{Lorenz1963,
	author = {{Lorenz}, Edward N.},
	title = "{Deterministic Nonperiodic Flow.}",
	journal = {Journal of the Atmospheric Sciences},
	year = 1963,
	month = mar,
	volume = {20},
	number = {2},
	pages = {130-148},
	doi = {10.1175/1520-0469(1963)020<0130:DNF>2.0.CO;2},
	adsurl = {https://ui.adsabs.harvard.edu/abs/1963JAtS...20..130L}
}

@BOOK{Valtonen2006,
	author = {{Valtonen}, Mauri and {Karttunen}, Hannu},
	title = "{The Three-Body Problem}",
	publisher = {Cambridge University Press},
	year = 2006,
	address = {Cambridge, UK},
	isbn = {978-0-521-85224-1},
	adsurl = {https://ui.adsabs.harvard.edu/abs/2006tbp..book.....V}
}

@ARTICLE{Fregeau2006,
	author = {{Fregeau}, John M. and {Rasio}, Frederic A.},
	title = "{Monte Carlo Simulations of Globular Cluster Evolution. IV. Direct Integration of Strong Interactions}",
	journal = {\apj},
	year = 2007,
	month = apr,
	volume = {658},
	number = {2},
	pages = {1047-1061},
	doi = {10.1086/511809},
	archivePrefix = {arXiv},
	eprint = {astro-ph/0608261},
	primaryClass = {astro-ph},
	adsurl = {https://ui.adsabs.harvard.edu/abs/2007ApJ...658.1047F}
}

@BOOK{Heggie2003,
	author = {{Heggie}, Douglas and {Hut}, Piet},
	title = "{The Gravitational Million-Body Problem: A Multidisciplinary Approach to Star Cluster Dynamics}",
	publisher = {Cambridge University Press},
	year = 2003,
	address = {Cambridge, UK},
	isbn = {978-0-521-77303-4},
	adsurl = {https://ui.adsabs.harvard.edu/abs/2003gmbp.book.....H}
}

@ARTICLE{Hut1983,
	author = {{Hut}, P. and {Bahcall}, J.~N.},
	title = "{Binary-single star scattering. I - Numerical experiments for equal masses}",
	journal = {\apj},
	year = 1983,
	month = may,
	volume = {268},
	pages = {319-341},
	doi = {10.1086/160956},
	adsurl = {https://ui.adsabs.harvard.edu/abs/1983ApJ...268..319H}
}

@ARTICLE{SiTu2025,
	author = {{SiTu}, Hengjian and {Wu}, Xiao-Jun and {Yang}, Bo and {Lin}, Wenbin},
	title = "{Self-attention-enhanced Deep Neural Networks for Simulating Post-Newtonian Dynamics of Three-body Systems}",
	journal = {\apjs},
	year = 2025,
	month = dec,
	volume = {281},
	number = {2},
	eid = {42},
	pages = {42},
	doi = {10.3847/1538-4365/ae173f},
	adsurl = {https://ui.adsabs.harvard.edu/abs/2025ApJS..281...42S}
}

@ARTICLE{Pereira2024,
	author = {{Santos Pereira}, Manuel and {Tripa}, Lu{\'\i}s and {Lima}, N{\'e}lson and {Caldas}, Francisco and {Soares}, Cl{\'a}udia},
	title = "{Advancing Solutions for the Three-Body Problem Through Physics-Informed Neural Networks}",
	journal = {arXiv e-prints},
	year = 2025,
	month = mar,
	eid = {arXiv:2503.04585},
	pages = {arXiv:2503.04585},
	doi = {10.48550/arXiv.2503.04585},
	archivePrefix = {arXiv},
	eprint = {2503.04585},
	primaryClass = {cs.LG},
	adsurl = {https://ui.adsabs.harvard.edu/abs/2025arXiv250304585S}
}

@ARTICLE{Carita2024,
	author = {{Carit{\'a}}, G.~A. and {Aljbaae}, S. and {Morais}, M.~H.~M. and {Signor}, A.~C. and {Carruba}, V. and {Prado}, A.~F.~B.~A. and {Hussmann}, H.},
	title = "{Image classification of retrograde resonance in the planar circular restricted three-body problem}",
	journal = {Celestial Mechanics and Dynamical Astronomy},
	year = 2024,
	month = apr,
	volume = {136},
	number = {2},
	eid = {10},
	pages = {10},
	doi = {10.1007/s10569-024-10181-8},
	adsurl = {https://ui.adsabs.harvard.edu/abs/2024CeMDA.136...10C}
}

@ARTICLE{Kroupa2001,
	author = {{Kroupa}, Pavel},
	title = "{On the variation of the initial mass function}",
	journal = {\mnras},
	year = 2001,
	month = apr,
	volume = {322},
	number = {2},
	pages = {231-246},
	doi = {10.1046/j.1365-8711.2001.04022.x},
	archivePrefix = {arXiv},
	eprint = {astro-ph/0009005},
	primaryClass = {astro-ph},
	adsurl = {https://ui.adsabs.harvard.edu/abs/2001MNRAS.322..231K}
}

@ARTICLE{Weidner2004,
	author = {{Weidner}, C. and {Kroupa}, P.},
	title = "{Evidence for a fundamental stellar upper mass limit from clustered star formation}",
	journal = {\mnras},
	year = 2004,
	month = feb,
	volume = {348},
	number = {1},
	pages = {187-191},
	doi = {10.1111/j.1365-2966.2004.07340.x},
	archivePrefix = {arXiv},
	eprint = {astro-ph/0310860},
	primaryClass = {astro-ph},
	adsurl = {https://ui.adsabs.harvard.edu/abs/2004MNRAS.348..187W}
}

@ARTICLE{Bojnordi2021,
	author = {{Bojnordi Arbab}, Behzad and {Rahvar}, Sohrab},
	title = "{Close stellar encounters kicking planets out of habitable zone in various stellar environments}",
	journal = {International Journal of Modern Physics D},
	year = 2021,
	month = jul,
	volume = {30},
	number = {9},
	eid = {2150063},
	pages = {2150063},
	doi = {10.1142/S0218271821500632},
	archivePrefix = {arXiv},
	eprint = {2001.10595},
	primaryClass = {astro-ph.EP},
	adsurl = {https://ui.adsabs.harvard.edu/abs/2021IJMPD..3050063B}
}

@BOOK{Binney2008,
	author = {{Binney}, James and {Tremaine}, Scott},
	title = "{Galactic Dynamics: Second Edition}",
	publisher = {Princeton University Press},
	year = 2008,
	address = {Princeton, NJ},
	isbn = {978-0-691-13027-9},
	adsurl = {https://ui.adsabs.harvard.edu/abs/2008gady.book.....B}
}

@ARTICLE{Torres2013,
	author = {{Jim{\'e}nez-Torres}, Juan J. and {Pichardo}, B{\'a}rbara and {Lake}, George and {Segura}, Ant{\'\i}gona},
	title = "{Habitability in Different Milky Way Stellar Environments: A Stellar Interaction Dynamical Approach}",
	journal = {Astrobiology},
	year = 2013,
	month = may,
	volume = {13},
	number = {5},
	pages = {491-509},
	doi = {10.1089/ast.2012.0842},
	archivePrefix = {arXiv},
	eprint = {1306.0464},
	primaryClass = {astro-ph.EP},
	adsurl = {https://ui.adsabs.harvard.edu/abs/2013AsBio..13..491J}
}

@ARTICLE{Rein2012,
	author = {{Rein}, H. and {Liu}, S.-F.},
	title = "{REBOUND: an open-source multi-purpose N-body code for collisional dynamics}",
	journal = {\aap},
	year = 2012,
	month = jan,
	volume = {537},
	eid = {A128},
	pages = {A128},
	doi = {10.1051/0004-6361/201118085},
	archivePrefix = {arXiv},
	eprint = {1110.4876},
	primaryClass = {astro-ph.EP},
	adsurl = {https://ui.adsabs.harvard.edu/abs/2012A&A...537A.128R}
}

@ARTICLE{Rein2015,
	author = {{Rein}, Hanno and {Spiegel}, David S.},
	title = "{IAS15: a fast, adaptive, high-order integrator for gravitational dynamics, accurate to machine precision over a billion orbits}",
	journal = {\mnras},
	year = 2015,
	month = jan,
	volume = {446},
	number = {2},
	pages = {1424-1437},
	doi = {10.1093/mnras/stu2164},
	archivePrefix = {arXiv},
	eprint = {1409.4779},
	primaryClass = {astro-ph.EP},
	adsurl = {https://ui.adsabs.harvard.edu/abs/2015MNRAS.446.1424R}
}

@INPROCEEDINGS{Davies1996,
	author = {{Davies}, M.~B.},
	title = "{Stellar Encounters in Dense Systems}",
	booktitle = {Dynamical Evolution of Star Clusters: Confrontation of Theory and Observations},
	year = 1996,
	editor = {{Hut}, Piet and {Makino}, Junichiro},
	series = {IAU Symposium},
	volume = {174},
	month = jan,
	pages = {243},
	adsurl = {https://ui.adsabs.harvard.edu/abs/1996IAUS..174..243D}
}

@ARTICLE{Trani2019,
	author = {{Trani}, Alessandro A. and {Spera}, Mario and {Leigh}, Nathan W.~C. and {Fujii}, Michiko S.},
	title = "{The Keplerian Three-body Encounter. II. Comparisons with Isolated Encounters and Impact on Gravitational Wave Merger Timescales}",
	journal = {\apj},
	year = 2019,
	month = nov,
	volume = {885},
	number = {2},
	eid = {135},
	pages = {135},
	doi = {10.3847/1538-4357/ab480a},
	archivePrefix = {arXiv},
	eprint = {1904.07879},
	primaryClass = {astro-ph.GA},
	adsurl = {https://ui.adsabs.harvard.edu/abs/2019ApJ...885..135T}
}

@INPROCEEDINGS{Heggie1996,
	author = {{Heggie}, D.~C. and {Hut}, P. and {McMillan}, S.~L.~W.},
	title = "{Exchange Cross Sections For Hard Binaries}",
	booktitle = {Dynamical Evolution of Star Clusters: Confrontation of Theory and Observations},
	year = 1996,
	editor = {{Hut}, Piet and {Makino}, Junichiro},
	series = {IAU Symposium},
	volume = {174},
	month = jan,
	pages = {371},
	doi = {10.48550/arXiv.astro-ph/9603080},
	archivePrefix = {arXiv},
	eprint = {astro-ph/9603080},
	primaryClass = {astro-ph},
	adsurl = {https://ui.adsabs.harvard.edu/abs/1996IAUS..174..371H}
}

@ARTICLE{Hall1996,
	author = {{Hall}, S.~M. and {Clarke}, C.~J. and {Pringle}, J.~E.},
	title = "{Energetics of star-disc encounters in the non-linear regime}",
	journal = {\mnras},
	year = 1996,
	month = jan,
	volume = {278},
	pages = {303-320},
	doi = {10.1093/mnras/278.2.303},
	archivePrefix = {arXiv},
	eprint = {astro-ph/9510153},
	primaryClass = {astro-ph},
	adsurl = {https://ui.adsabs.harvard.edu/abs/1996MNRAS.278..303H}
}

@ARTICLE{Pedregosa2011,
	author = {{Pedregosa}, Fabian and {Varoquaux}, Ga{\"e}l and {Gramfort}, Alexandre and {Michel}, Vincent and {Thirion}, Bertrand and {Grisel}, Olivier and {Blondel}, Mathieu and {M{\"u}ller}, Andreas and {Nothman}, Joel and {Louppe}, Gilles and {Prettenhofer}, Peter and {Weiss}, Ron and {Dubourg}, Vincent and {Vanderplas}, Jake and {Passos}, Alexandre and {Cournapeau}, David and {Brucher}, Matthieu and {Perrot}, Matthieu and {Duchesnay}, {\'E}douard},
	title = "{Scikit-learn: Machine Learning in Python}",
	journal = {Journal of Machine Learning Research},
	year = 2011,
	month = oct,
	volume = {12},
	pages = {2825-2830},
	doi = {10.48550/arXiv.1201.0490},
	archivePrefix = {arXiv},
	eprint = {1201.0490},
	primaryClass = {cs.LG},
	adsurl = {https://ui.adsabs.harvard.edu/abs/2011JMLR...12.2825P}
}

@ARTICLE{Pinheiro2025,
	author = {{Manoel Herrera Pinheiro}, Jo{\~a}o and {Vilas Boas de Oliveira}, Suzana and {Segreto Silva}, Thiago Henrique and {Rabelo Saraiva}, Pedro Antonio and {Ferreira de Souza}, Enzo and {Godoy}, Ricardo V. and {Andr{\'e} Ambrosio}, Leonardo and {Becker}, Marcelo},
	title = "{The Impact of Feature Scaling In Machine Learning: Effects on Regression and Classification Tasks}",
	journal = {arXiv e-prints},
	year = 2025,
	month = jun,
	eid = {arXiv:2506.08274},
	pages = {arXiv:2506.08274},
	doi = {10.48550/arXiv.2506.08274},
	archivePrefix = {arXiv},
	eprint = {2506.08274},
	primaryClass = {stat.ML},
	adsurl = {https://ui.adsabs.harvard.edu/abs/2025arXiv250608274M}
}

@ARTICLE{Chen2016,
	author = {{Chen}, Tianqi and {Guestrin}, Carlos},
	title = "{XGBoost: A Scalable Tree Boosting System}",
	journal = {arXiv e-prints},
	year = 2016,
	month = mar,
	eid = {arXiv:1603.02754},
	pages = {arXiv:1603.02754},
	doi = {10.48550/arXiv.1603.02754},
	archivePrefix = {arXiv},
	eprint = {1603.02754},
	primaryClass = {cs.LG},
	adsurl = {https://ui.adsabs.harvard.edu/abs/2016arXiv160302754C}
}

@mastersthesis{Albertyn2025,
	author = {Albertyn, C.},
	title = {A comparative study of gradient boosting algorithms and transformer neural networks for classification problems in the tabular data domain},
	school = {Stellenbosch University},
	year = {2025},
	address = {Stellenbosch, South Africa},
	month = {March},
	note = {MEng Thesis},
	url = {https://scholar.sun.ac.za/items/1426022c-fbf0-44eb-ab74-43df924264d6}
}

@ARTICLE{Mojiri2021,
	author = {{Mojiri}, Arezou and {Khalili}, Abbas and {Zeinal Hamadani}, Ali},
	title = "{New Hard-thresholding Rules based on Data Splitting in High-dimensional Imbalanced Classification}",
	journal = {arXiv e-prints},
	year = 2021,
	month = nov,
	eid = {arXiv:2111.03306},
	pages = {arXiv:2111.03306},
	doi = {10.48550/arXiv.2111.03306},
	archivePrefix = {arXiv},
	eprint = {2111.03306},
	primaryClass = {stat.ME},
	adsurl = {https://ui.adsabs.harvard.edu/abs/2021arXiv211103306M}
}

@ARTICLE{Pasquato2024,
	author = {{Pasquato}, Mario and {Trevisan}, Piero and {Askar}, Abbas and {Lemos}, Pablo and {Carenini}, Gaia and {Mapelli}, Michela and {Hezaveh}, Yashar},
	title = "{Interpretable Machine Learning for Finding Intermediate-mass Black Holes}",
	journal = {\apj},
	year = 2024,
	month = apr,
	volume = {965},
	number = {1},
	eid = {89},
	pages = {89},
	doi = {10.3847/1538-4357/ad2261},
	archivePrefix = {arXiv},
	eprint = {2310.18560},
	primaryClass = {astro-ph.GA},
	adsurl = {https://ui.adsabs.harvard.edu/abs/2024ApJ...965...89P}
}

@ARTICLE{Valle2021,
	author = {{Machado Poletti Valle}, Luis Fernando and {Avestruz}, Camille and {Barnes}, David J. and {Farahi}, Arya and {Lau}, Erwin T. and {Nagai}, Daisuke},
	title = "{SHAPing the gas: understanding gas shapes in dark matter haloes with interpretable machine learning}",
	journal = {\mnras},
	year = 2021,
	month = oct,
	volume = {507},
	number = {1},
	pages = {1468-1484},
	doi = {10.1093/mnras/stab2252},
	archivePrefix = {arXiv},
	eprint = {2011.12987},
	primaryClass = {astro-ph.CO},
	adsurl = {https://ui.adsabs.harvard.edu/abs/2021MNRAS.507.1468M}
}

@ARTICLE{Aarseth1999,
	author = {{Aarseth}, Sverre J.},
	title = "{From NBODY1 to NBODY6: The Growth of an Industry}",
	journal = {\pasp},
	year = 1999,
	month = nov,
	volume = {111},
	number = {765},
	pages = {1333-1346},
	doi = {10.1086/316455},
	adsurl = {https://ui.adsabs.harvard.edu/abs/1999PASP..111.1333A}
}

@ARTICLE{Nawaz2025,
	author = {{Nawaz}, Ali and {Ahmad}, Amir and {Khan}, Shehroz S.},
	title = "{Beyond Rebalancing: Benchmarking Binary Classifiers Under Class Imbalance Without Rebalancing Techniques}",
	journal = {arXiv e-prints},
	year = 2025,
	month = sep,
	eid = {arXiv:2509.07605},
	pages = {arXiv:2509.07605},
	doi = {10.48550/arXiv.2509.07605},
	archivePrefix = {arXiv},
	eprint = {2509.07605},
	primaryClass = {stat.ML},
	adsurl = {https://ui.adsabs.harvard.edu/abs/2025arXiv250907605N}
}

@ARTICLE{Yokota2012,
	author = {{Yokota}, Rio and {Barba}, Lorena A.},
	title = "{Hierarchical N-Body Simulations with Autotuning for Heterogeneous Systems}",
	journal = {Computing in Science and Engineering},
	year = 2012,
	month = may,
	volume = {14},
	number = {3},
	pages = {30-39},
	doi = {10.1109/MCSE.2012.1},
	archivePrefix = {arXiv},
	eprint = {1108.5815},
	primaryClass = {cs.NA},
	adsurl = {https://ui.adsabs.harvard.edu/abs/2012CSE....14c..30Y}
}

@ARTICLE{Zwart2007,
	author = {{Portegies Zwart}, Simon and {Belleman}, Robert and {Geldof}, Peter},
	title = "{High Performance Direct Gravitational N-body Simulations on Graphics Processing Units}",
	journal = {arXiv e-prints},
	year = 2007,
	month = feb,
	eid = {cs/0702135},
	pages = {cs/0702135},
	doi = {10.48550/arXiv.cs/0702135},
	archivePrefix = {arXiv},
	eprint = {cs/0702135},
	primaryClass = {cs.PF},
	adsurl = {https://ui.adsabs.harvard.edu/abs/2007cs........2135P}
}

@book{Hairer1993,
	title={Solving Ordinary Differential Equations II: Stiff and Differential-Algebraic Problems},
	author={Hairer, E. and N{\o}rsett, S.P. and Wanner, G.},
	isbn={9783540604525},
	lccn={86031456},
	series={Solving Ordinary Differential Equations II: Stiff and Differential-algebraic Problems},
	url={https://books.google.com/books?id=m7c8nNLPwaIC},
	year={1993},
	publisher={Springer}
}

@ARTICLE{Zwart2014,
	author = {{Portegies Zwart}, Simon and {Boekholt}, Tjarda},
	title = "{On the Minimal Accuracy Required for Simulating Self-gravitating Systems by Means of Direct N-body Methods}",
	journal = {\apjl},
	year = 2014,
	month = apr,
	volume = {785},
	number = {1},
	eid = {L3},
	pages = {L3},
	doi = {10.1088/2041-8205/785/1/L3},
	archivePrefix = {arXiv},
	eprint = {1402.6713},
	primaryClass = {astro-ph.IM},
	adsurl = {https://ui.adsabs.harvard.edu/abs/2014ApJ...785L...3P}
}

@ARTICLE{Wang2020,
	author = {{Wang}, Long and {Iwasawa}, Masaki and {Nitadori}, Keigo and {Makino}, Junichiro},
	title = "{PETAR: a high-performance N-body code for modelling massive collisional stellar systems}",
	journal = {\mnras},
	year = 2020,
	month = sep,
	volume = {497},
	number = {1},
	pages = {536-555},
	doi = {10.1093/mnras/staa1915},
	archivePrefix = {arXiv},
	eprint = {2006.16560},
	primaryClass = {astro-ph.IM},
	adsurl = {https://ui.adsabs.harvard.edu/abs/2020MNRAS.497..536W}
}

@ARTICLE{Makarov2025,
	author = {{Makarov}, Valeri V.},
	title = "{Distributions of Wide Binary Stars in Theory and in Gaia Data. II. Reconstruction of Sample Probability Density of True Orbit Sizes}",
	journal = {\aj},
	year = 2025,
	month = sep,
	volume = {170},
	number = {3},
	eid = {138},
	pages = {138},
	doi = {10.3847/1538-3881/adec79},
	archivePrefix = {arXiv},
	eprint = {2507.19273},
	primaryClass = {astro-ph.SR},
	adsurl = {https://ui.adsabs.harvard.edu/abs/2025AJ....170..138M}
}

@article{Gholamy2018,
	author = {Gholamy, Afshin and Kreinovich, Vladik and Kosheleva, Olga},
	title = {Why 70/30 or 80/20 Relation Between Training and Testing Sets: A Pedagogical Explanation},
	journal = {Journal of Intelligent Technologies and Applied Statistics},
	volume = {11},
	number = {2},
	pages = {105--111},
	year = {2018}
}

@ARTICLE{Lin2020,
	author = {{Lin}, Haitao and {Li}, Xiangru and {Luo}, Ziying},
	title = "{Pulsars detection by machine learning with very few features}",
	journal = {\mnras},
	year = 2020,
	month = apr,
	volume = {493},
	number = {2},
	pages = {1842-1854},
	doi = {10.1093/mnras/staa218},
	archivePrefix = {arXiv},
	eprint = {2002.08519},
	primaryClass = {astro-ph.IM},
	adsurl = {https://ui.adsabs.harvard.edu/abs/2020MNRAS.493.1842L}
}

@ARTICLE{Alegre2022,
	author = {{Alegre}, Lara and {Sabater}, Jose and {Best}, Philip and {Mostert}, Rafa{\"e}l I.~J. and {Williams}, Wendy L. and {G{\"u}rkan}, G{\"u}lay and {Hardcastle}, Martin J. and {Kondapally}, Rohit and {Shimwell}, Tim W. and {Smith}, Daniel J.~B.},
	title = "{A machine-learning classifier for LOFAR radio galaxy cross-matching techniques}",
	journal = {\mnras},
	year = 2022,
	month = nov,
	volume = {516},
	number = {4},
	pages = {4716-4738},
	doi = {10.1093/mnras/stac1888},
	archivePrefix = {arXiv},
	eprint = {2207.01645},
	primaryClass = {astro-ph.IM},
	adsurl = {https://ui.adsabs.harvard.edu/abs/2022MNRAS.516.4716A}
}

@techreport{Chen2004,
	author = {Chen, Chao and Liaw, Andy and Breiman, Leo},
	editor = {University of California, Berkeley 110 (1-12): 24},
	institution = {University of California, Berkeley 110 (1-12): 24},
	number = 666,
	title = {Using Random Forest to Learn Imbalanced Data},
	url = {https://statistics.berkeley.edu/sites/default/files/tech-reports/666.pdf},
	year = 2004
}

@ARTICLE{Wainer2018,
	author = {{Wainer}, Jacques},
	title = "{An empirical evaluation of imbalanced data strategies from a practitioner's point of view}",
	journal = {arXiv e-prints},
	year = 2018,
	month = oct,
	eid = {arXiv:1810.07168},
	pages = {arXiv:1810.07168},
	doi = {10.48550/arXiv.1810.07168},
	archivePrefix = {arXiv},
	eprint = {1810.07168},
	primaryClass = {cs.LG},
	adsurl = {https://ui.adsabs.harvard.edu/abs/2018arXiv181007168W}
}

@article{Sasirekha2025,
	author = {Sasirekha, R. and Kanisha, B.},
	title = {Adaptive Ensemble Framework With Synthetic Sampling for Tackling Class Imbalance Problem},
	journal = {Engineering Reports},
	volume = {7},
	number = {4},
	pages = {e70109},
	doi = {https://doi.org/10.1002/eng2.70109},
	url = {https://onlinelibrary.wiley.com/doi/abs/10.1002/eng2.70109},
	eprint = {https://onlinelibrary.wiley.com/doi/pdf/10.1002/eng2.70109},
	year = {2025}
}

@ARTICLE{Mortari2007,
	author = {{Mortari}, Daniele and {Clocchiatti}, Alberto},
	title = "{Solving Kepler's Equation using B{\'e}zier curves}",
	journal = {Celestial Mechanics and Dynamical Astronomy},
	year = 2007,
	month = sep,
	volume = {99},
	number = {1},
	pages = {45-57},
	doi = {10.1007/s10569-007-9089-2},
	adsurl = {https://ui.adsabs.harvard.edu/abs/2007CeMDA..99...45M}
}

@article{Bergstra2012,
	author  = {James Bergstra and Yoshua Bengio},
	title   = {Random Search for Hyper-Parameter Optimization},
	journal = {Journal of Machine Learning Research},
	year    = {2012},
	volume  = {13},
	number  = {10},
	pages   = {281--305},
	url     = {http://jmlr.org/papers/v13/bergstra12a.html}
}

@ARTICLE{Cesare2021,
	author = {{De Cesare}, G. and {Capuzzo-Dolcetta}, R.},
	title = "{On the stability of planetary orbits in binary star systems I. The S-type orbits}",
	journal = {\apss},
	year = 2021,
	month = jun,
	volume = {366},
	number = {6},
	eid = {53},
	pages = {53},
	doi = {10.1007/s10509-021-03959-x},
	archivePrefix = {arXiv},
	eprint = {2106.01753},
	primaryClass = {astro-ph.EP},
	adsurl = {https://ui.adsabs.harvard.edu/abs/2021Ap&SS.366...53D}
}

@ARTICLE{Tory2022,
	author = {{Tory}, Max and {Grishin}, Evgeni and {Mandel}, Ilya},
	title = "{Empirical stability boundary for hierarchical triples}",
	journal = {\pasa},
	year = 2022,
	month = nov,
	volume = {39},
	eid = {e062},
	pages = {e062},
	doi = {10.1017/pasa.2022.57},
	archivePrefix = {arXiv},
	eprint = {2208.14005},
	primaryClass = {astro-ph.SR},
	adsurl = {https://ui.adsabs.harvard.edu/abs/2022PASA...39...62T}
}

@ARTICLE{Dolcetta2020,
	author = {{Capuzzo-Dolcetta}, Roberto and {De Cesare}, Giovanni and {Marino}, Alessio},
	title = "{Stability of Planetary Motion in Binary Star Systems}",
	journal = {arXiv e-prints},
	year = 2020,
	month = aug,
	eid = {arXiv:2008.11107},
	pages = {arXiv:2008.11107},
	doi = {10.48550/arXiv.2008.11107},
	archivePrefix = {arXiv},
	eprint = {2008.11107},
	primaryClass = {astro-ph.EP},
	adsurl = {https://ui.adsabs.harvard.edu/abs/2020arXiv200811107C}
}

@ARTICLE{Wisdom1991,
	author = {{Wisdom}, Jack and {Holman}, Matthew},
	title = "{Symplectic maps for the N-body problem.}",
	journal = {\aj},
	year = 1991,
	month = oct,
	volume = {102},
	pages = {1528-1538},
	doi = {10.1086/115978},
	adsurl = {https://ui.adsabs.harvard.edu/abs/1991AJ....102.1528W}
}

@ARTICLE{Rodriguez2022,
	author = {{Rodriguez}, Carl L. and {Weatherford}, Newlin C. and {Coughlin}, Scott C. and {Amaro-Seoane}, Pau and {Breivik}, Katelyn and {Chatterjee}, Sourav and {Fragione}, Giacomo and {K{\i}ro{\u{g}}lu}, Fulya and {Kremer}, Kyle and {Rui}, Nicholas Z. and {Ye}, Claire S. and {Zevin}, Michael and {Rasio}, Frederic A.},
	title = "{Modeling Dense Star Clusters in the Milky Way and beyond with the Cluster Monte Carlo Code}",
	journal = {\apjs},
	year = 2022,
	month = feb,
	volume = {258},
	number = {2},
	eid = {22},
	pages = {22},
	doi = {10.3847/1538-4365/ac2edf},
	archivePrefix = {arXiv},
	eprint = {2106.02643},
	primaryClass = {astro-ph.GA},
	adsurl = {https://ui.adsabs.harvard.edu/abs/2022ApJS..258...22R}
}

@ARTICLE{Atallah2025,
	author = {{Atallah}, Dany and {Ginat}, Yonadav Barry and {Weatherford}, Newlin C.},
	title = "{A Million Three-body Binaries Caught by Gaia}",
	journal = {\apj},
	year = 2025,
	month = nov,
	volume = {993},
	number = {2},
	eid = {183},
	pages = {183},
	doi = {10.3847/1538-4357/ae073c},
	archivePrefix = {arXiv},
	eprint = {2503.14605},
	primaryClass = {astro-ph.GA},
	adsurl = {https://ui.adsabs.harvard.edu/abs/2025ApJ...993..183A}
}

@ARTICLE{Samsing2018,
	author = {{Samsing}, Johan and {MacLeod}, Morgan and {Ramirez-Ruiz}, Enrico},
	title = "{Dissipative Evolution of Unequal-mass Binary-single Interactions and Its Relevance to Gravitational-wave Detections}",
	journal = {\apj},
	year = 2018,
	month = feb,
	volume = {853},
	number = {2},
	eid = {140},
	pages = {140},
	doi = {10.3847/1538-4357/aaa715},
	archivePrefix = {arXiv},
	eprint = {1706.03776},
	primaryClass = {astro-ph.HE},
	adsurl = {https://ui.adsabs.harvard.edu/abs/2018ApJ...853..140S}
}

@ARTICLE{Marks2012,
	author = {{Marks}, M. and {Kroupa}, P.},
	title = "{Inverse dynamical population synthesis. Constraining the initial conditions of young stellar clusters by studying their binary populations}",
	journal = {\aap},
	year = 2012,
	month = jul,
	volume = {543},
	eid = {A8},
	pages = {A8},
	doi = {10.1051/0004-6361/201118231},
	archivePrefix = {arXiv},
	eprint = {1205.1508},
	primaryClass = {astro-ph.GA},
	adsurl = {https://ui.adsabs.harvard.edu/abs/2012A&A...543A...8M}
}

@ARTICLE{Ivanova2005,
	author = {{Ivanova}, N. and {Belczynski}, K. and {Fregeau}, J.~M. and {Rasio}, F.~A.},
	title = "{The evolution of binary fractions in globular clusters}",
	journal = {\mnras},
	year = 2005,
	month = apr,
	volume = {358},
	number = {2},
	pages = {572-584},
	doi = {10.1111/j.1365-2966.2005.08804.x},
	archivePrefix = {arXiv},
	eprint = {astro-ph/0501131},
	primaryClass = {astro-ph},
	adsurl = {https://ui.adsabs.harvard.edu/abs/2005MNRAS.358..572I}
}

@ARTICLE{Stacy2010,
	author = {{Stacy}, Athena and {Greif}, Thomas H. and {Bromm}, Volker},
	title = "{The first stars: formation of binaries and small multiple systems}",
	journal = {\mnras},
	year = 2010,
	month = mar,
	volume = {403},
	number = {1},
	pages = {45-60},
	doi = {10.1111/j.1365-2966.2009.16113.x},
	archivePrefix = {arXiv},
	eprint = {0908.0712},
	primaryClass = {astro-ph.CO},
	adsurl = {https://ui.adsabs.harvard.edu/abs/2010MNRAS.403...45S}
}

@ARTICLE{Hamilton2024,
	author = {{Hamilton}, Chris and {Modak}, Shaunak},
	title = "{Eccentricity dynamics of wide binaries - II. The effect of stellar encounters and constraints on formation channels}",
	journal = {\mnras},
	year = 2024,
	month = aug,
	volume = {532},
	number = {2},
	pages = {2425-2440},
	doi = {10.1093/mnras/stae1654},
	archivePrefix = {arXiv},
	eprint = {2311.04352},
	primaryClass = {astro-ph.GA},
	adsurl = {https://ui.adsabs.harvard.edu/abs/2024MNRAS.532.2425H}
}

@ARTICLE{Mathew2024,
	author = {{Mathew}, Sajay Sunny and {Xu}, Siyao and {Federrath}, Christoph and {Hu}, Yue and {Seta}, Amit},
	title = "{Wide-binary eccentricity distribution in young star clusters: dependence on the binary separation and mass}",
	journal = {\mnras},
	year = 2024,
	month = aug,
	volume = {532},
	number = {2},
	pages = {2374-2387},
	doi = {10.1093/mnras/stae1632},
	archivePrefix = {arXiv},
	eprint = {2406.18184},
	primaryClass = {astro-ph.GA},
	adsurl = {https://ui.adsabs.harvard.edu/abs/2024MNRAS.532.2374M}
}
	
\clearpage
	
\begin{appendix}

\nolinenumbers
		
	\section{Orbital Parameter Distributions}
	\label{app:orbital}
		
\begin{figure}[h!]
		\centering
		\includegraphics[width=\columnwidth]{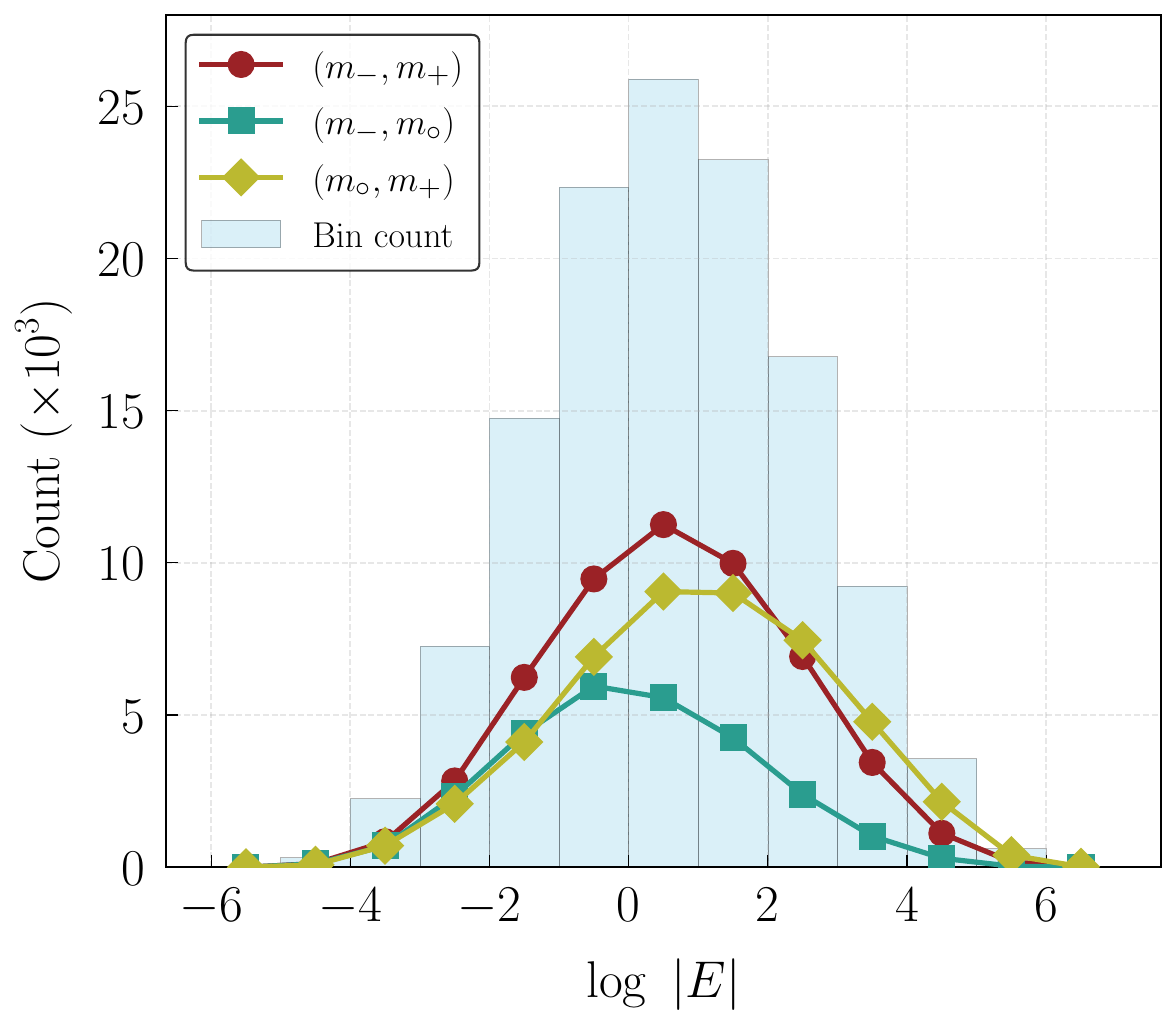}
		\caption{Binding energy distribution of the binaries formed in simulated three-body encounters. Colored lines show the number of binaries for each mass-rank pairing of the binary components.}
		\label{fig:energy_mass_pairs}
	\end{figure}

	Figure~\ref{fig:energy_mass_pairs} shows the distribution of the binaries formed in our simulations as a function of their binding energy $|E|$. The light blue histogram represents the overall binary population. The colored lines show the mass-rank composition of the formed binaries, where $m_-$, $m_\circ$, and $m_+$ denote the least, intermediate, and most massive stars in each encounter, respectively. The red, green, and yellow lines correspond to binaries composed of ($m_-$, $m_+$), ($m_-$, $m_\circ$), and ($m_\circ$, $m_+$), respectively.  
		
	Our findings show a different mass-pairing preference than that reported by \citet{Atallah2024}. While they found that wide binaries preferentially form from the two least massive bodies and hard binaries from the most and least massive bodies, our results indicate the opposite: wide binaries favor pairing the least and most massive bodies, whereas tight binaries preferentially form from the two most massive bodies (see Figure~\ref{fig:energy_mass_pairs}). This discrepancy may arise from differences in the physical assumptions and parameter sampling between the two studies. First, our simulations treat stars as point masses evolving under purely Newtonian gravity, whereas \citet{Atallah2024} included finite-size effects and post-Newtonian corrections. Second, the initial parameter distributions differ substantially: we sampled masses log-uniformly over a continuous range of $[0.08, 150]\,M_\odot$ and velocities log-uniformly over $[0.01, 100]$ au yr$^{-1}$ ($\sim0.05$--$500$ km s$^{-1}$), while \citet{Atallah2024} adopted six discrete mass values and a broader velocity range spanning $[1, 3000]$ km s$^{-1}$.
		
	\begin{table}[b!]
		\centering
		\caption{Summary statistics of orbital parameters for the binaries in our cleaned dataset. Listed are the semimajor axis $a$ in au, eccentricity $e$, inclination $i$ in degrees, and mass ratio $q = m_2/m_1$ ($m_1 \geq m_2$). The columns report the mean, median, standard deviation, minimum, and maximum values for each parameter.}
		\begin{tabular}{lcccccc}
			\toprule
			& Mean & Median & Std & Min & Max \\
			\midrule
			$a$ & $3017$ & $74.13$ & $9670$ & $9.5\times 10^{-4}$ & $10^5$ \\
			$e$ & $0.85$ & $0.93$ & $0.20$ & $6.5\times 10^{-3}$ & $0.9999$ \\
			$i$ & $90.0$ & $90.0$ & $39.1$ & $2.1\times 10^{-1}$ & $180$ \\
			$q$ & $0.33$ & $0.26$ & $0.28$ & $5.8\times 10^{-4}$ & $1.0$ \\
			\bottomrule
		\end{tabular}
		\label{tab:orbital_stats}
	\end{table}
		
	The statistics in Table~\ref{tab:orbital_stats} characterize the orbital properties of the formed binaries. The semimajor axis distribution spans 9 orders of magnitude, from $9.5\times10^{-4}$ au to $10^5$ au, with a median of $74.13$ au and a mean of $3017$ au, indicating a heavy-tailed distribution dominated by wide binaries. The eccentricity distribution is strongly superthermal, with a mean and median of $0.85$ and $0.93$, respectively. This is consistent with the findings of \citet{Atallah2024}, who showed that three-body binary formation overwhelmingly favors wide binaries with superthermal eccentricities. Such superthermal distributions have also been observed in Gaia wide binaries \citep{Hamilton2024, Mathew2024}, and theoretical work has shown that the eccentricity distribution of three-body-formed binaries is generically superthermal for soft binaries \citep{Ginat2024}. The inclination distribution is approximately isotropic, with a mean of $90.0^\circ$ and a near-uniform spread between $0^\circ$ and $180^\circ$, as expected for randomly oriented orbits. The mass ratio distribution peaks at low values (median $q=0.26$), reflecting the predominance of highly unequal-mass pairings. This is expected from our finding that binaries preferentially form from the least and most massive bodies.
		
	\section{Sample Encounter Trajectories}
	\label{app:sample_visual}
		
\begin{figure*}
		\centering
		\includegraphics[width=\linewidth]{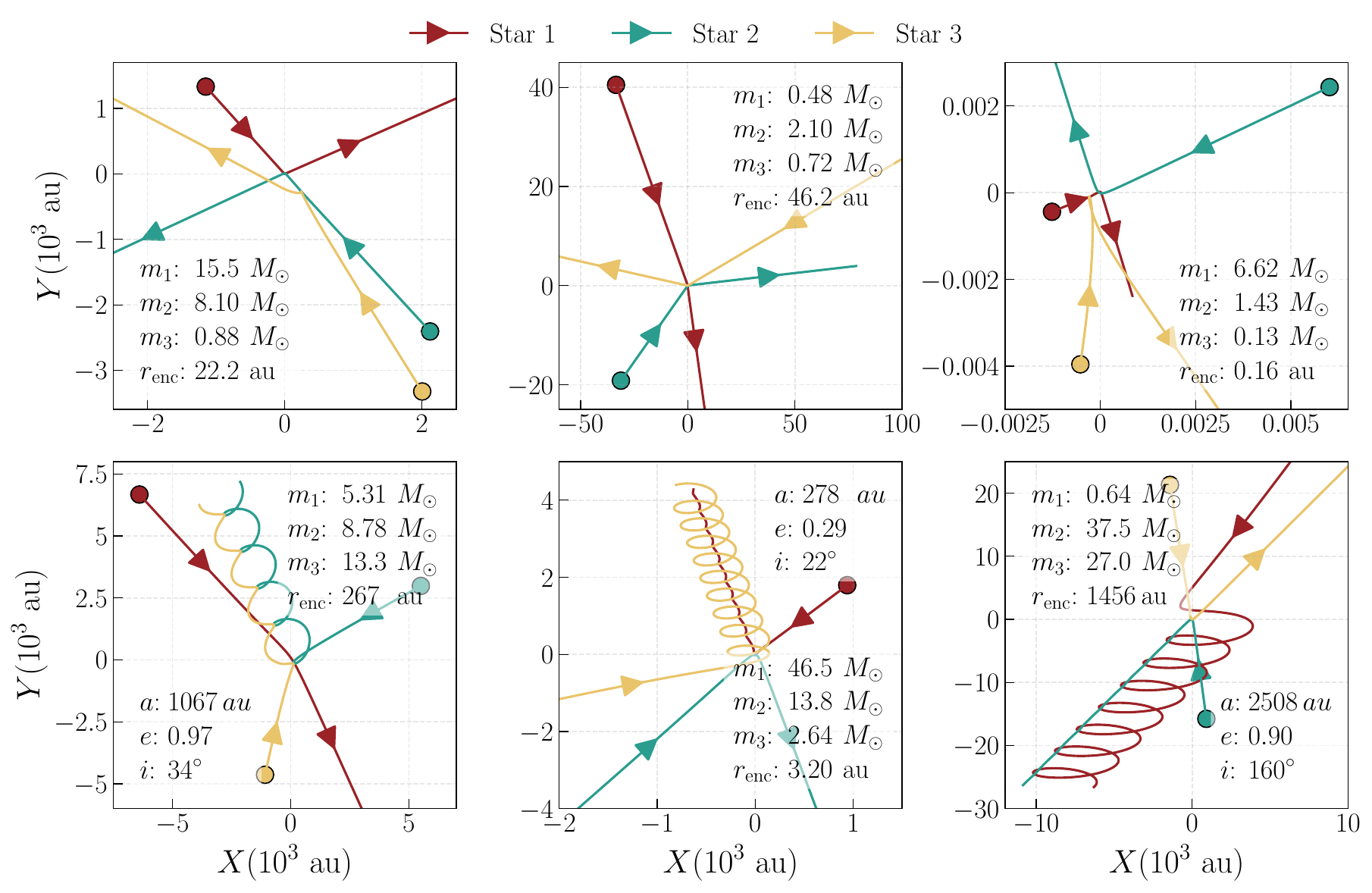}
		\caption{Example trajectories of three-body encounters.}
		\label{fig:trajectories_comparison}
	\end{figure*}

	Figure~\ref{fig:trajectories_comparison} shows example trajectories of six three-body encounters from our simulation campaign, projected onto the $x$-$y$ plane. The top panel illustrates fly-by encounters, in which no binary forms and all three stars remain unbound, even though they exchange momentum and energy through gravitational interactions. The bottom panel shows binary formation events, where two stars become gravitationally bound while the third escapes, carrying away the excess energy and angular momentum. Start points are marked with circles, and arrows indicate the motion directions. The masses are color-coded in red, green, and yellow for stars 1, 2, and 3, respectively. The encounter radius $r_{\mathrm{enc}}$ is indicated for all cases; for the binary formation events, the semi-major axis $a$, eccentricity $e$, and orbital inclination $i$ of the formed binary are also shown.
		
\end{appendix}
	
\end{document}